\documentclass[accepted]{melba}
\pdfoutput=1
\usepackage{mwe} 

\usepackage{amsmath,amsfonts}

\melbaid{2025:005}  
\doi{10.59275/j.melba.2025-f6fg}
\melbaauthors{Fathi Kazerooni, et al.}  
\email{Anahita.Kazerooni@pennmedicine.upenn.edu}
\volume{3}
\firstpageno{72}  
\melbayear{2025}  
\datesubmitted{2024-7-19}  
\datepublished{2025-5-28}  

\melbaspecialissue{Medical Imaging with Deep Learning (MIDL) 2020}
\melbaspecialissueeditors{Marleen de Bruijne, Tal Arbel, Ismail Ben Ayed, Hervé Lombaert}

\ShortHeadings{BraTS-PEDs Results 2023}{Fathi Kazerooni, et al.}

\title{BraTS-PEDs: Results of the Multi-Consortium International Pediatric Brain Tumor Segmentation Challenge 2023}

\author{
	\firstname Anahita \surname Fathi Kazerooni\aff{1},
    \firstname Nastaran \surname Khalili\aff{1},
    \firstname Xinyang \surname Liu\aff{2},
    \firstname Debanjan \surname Haldar\aff{3},
    \firstname Zhifan \surname Jiang\aff{2},
    \firstname Anna \surname Zapaishchykova\aff{4},
    \firstname Julija \surname Pavaine\aff{5},
    \firstname Lubdha M.  \surname Shah\aff{6},
    \firstname Blaise V. \surname Jones\aff{7},
    \firstname Nakul \surname Sheth\aff{8},
    \firstname Sanjay P. \surname Prabhu\aff{9},
    \firstname Aaron S. \surname McAllister\aff{10},
    \firstname Wenxin  \surname Tu\aff{11},
    \firstname Khanak K. \surname Nandolia\aff{12},
    \firstname Andres F.  \surname Rodriguez\aff{13},
    \firstname Ibraheem Salman  \surname Shaikh\aff{14},
    \firstname Mariana  \surname Sanchez-Montano\aff{15},
    \firstname Hollie Anne  \surname Lai\aff{16},
    \firstname Maruf \surname Adewole\aff{17},
    \firstname Jake \surname Albrecht\aff{18},
    \firstname Udunna \surname Anazodo\aff{19},
    \firstname Hannah \surname Anderson\aff{20},
    \firstname Syed Muhammed \surname Anwar\aff{2},
    \firstname Alejandro \surname Aristizabal\aff{21},
    \firstname Sina \surname Bagheri\aff{20},
    \firstname Ujjwal \surname Baid\aff{22},
    \firstname Timothy \surname Bergquist\aff{18},
    \firstname Austin J. \surname Borja\aff{23},
    \firstname Evan \surname Calabrese\aff{24},
    \firstname Verena \surname Chung\aff{18},
    \firstname Gian-Marco \surname Conte\aff{25},
    \firstname James \surname Eddy\aff{18},
    \firstname Ivan \surname Ezhov\aff{26},
    \firstname Ariana M. \surname Familiar\aff{1},
    \firstname Keyvan \surname Farahani\aff{27},
    \firstname Deep \surname Gandhi\aff{1},
    \firstname Anurag \surname Gottipati\aff{1},
    \firstname Shuvanjan \surname Haldar\aff{28},
    \firstname Juan Eugenio \surname Iglesias\aff{29},
    \firstname Anastasia \surname Janas\aff{30},
    \firstname Elaine \surname Elaine\aff{31},
    \firstname Alexandros \surname Karargyris\aff{21},
    \firstname Hasan \surname Kassem\aff{21},
    \firstname Neda \surname Khalili\aff{1},
    \firstname Florian \surname Kofler\aff{32},
    \firstname Dominic \surname LaBella\aff{33},
    \firstname Koen \surname Van Leemput\aff{34},
    \firstname Hongwei B. \surname Li\aff{35},
    \firstname Nazanin \surname Maleki\aff{30},
    \firstname Zeke \surname Meier\aff{36},
    \firstname Bjoern \surname Menze\aff{37},
    \firstname Ahmed W. \surname Moawad\aff{38},
    \firstname Sarthak \surname Pati\aff{21},
    \firstname Marie \surname Piraud\aff{32},
    \firstname Tina \surname Poussaint\aff{4},
    \firstname Zachary J. \surname Reitman\aff{33},
    \firstname Jeffrey D. \surname Rudie\aff{39},
    \firstname Rachit \surname Saluja\aff{40},
    \firstname MIcah \surname Sheller\aff{21},
    \firstname Russell Takeshi \surname Shinohara\aff{41},
    \firstname Karthik \surname Viswanathan\aff{1},
    \firstname Chunhao \surname Wang\aff{33},
    \firstname Benedikt \surname Wiestler\aff{42},
    \firstname Walter F. \surname Wiggins\aff{43},
    \firstname Christos \surname Davatzikos\aff{44},
    \firstname Phillip B. \surname Storm\aff{1},
    \firstname Miriam \surname Bornhorst\aff{45},
    \firstname Roger \surname Packer\aff{45},
    \firstname Trent  \surname Hummel\aff{46},
    \firstname Peter \surname de Blank\aff{46},
    \firstname Lindsey \surname Hoffman\aff{47},
    \firstname Mariam \surname Aboian\aff{8},
    \firstname Ali \surname Nabavizadeh\aff{1},
    \firstname Jeffrey B. \surname Ware\aff{1},
    \firstname Benjamin H. \surname Kann\aff{4},
    \firstname Brian \surname Rood\aff{48},
    \firstname Adam \surname Resnick\aff{1},
    \firstname Spyridon \surname Bakas\aff{22},
    \firstname Arastoo \surname Vossough\aff{8},
    \firstname Marius George  \surname Linguraru\aff{2}
}
\affiliations{
	\num 1 \addr Center for Data-Driven Discovery in Biomedicine (D3b), The Children's Hospital of Philadelphia, PA, USA,
	\num 2 \addr Sheikh Zayed Institute for Pediatric Surgical Innovation, Children's National Hospital, Washington DC, USA,
    \num 3 \addr Department of Neurosurgery, Thomas Jefferson University Hospital, PA, USA,
    \num 4 \addr Dana-Farber Brigham Cancer Center \& Boston Children’s Hospital, Boston, MA, USA,
    \num 5 \addr Department of Paediatric Radiology, Royal Manchester Children’s Hospital, Manchester University Hospitals NHS Foundation Trust, University of Manchester, Manchester, UK,
    \num 6 \addr University of Utah, UT, USA,
    \num 7 \addr Cincinnati Children’s Hospital Medical Center, OH, USA,
    \num 8 \addr Department of Radiology, The Children's Hospital of Philadelphia, PA, USA,
    \num 9 \addr Division of Neuroradiology, Department of Radiology, Boston Children's Hospital, Harvard Medical School, Boston, MA, USA,
    \num 10 \addr Department of Radiology, Nationwide Children's Hospital, OH, USA,
    \num 11 \addr College of Arts and Sciences, University of Pennsylvania, PA, USA,
    \num 12 \addr Department of Diagnostic and Interventional Radiology, All India Institute of Medical Sciences, Rishikesh, India,
    \num 13 \addr Department of Diagnostic \& Interventional Imaging, Neuroradiology section, The University of Texas Health at Houston, Texas, USA,
    \num 14 \addr Beth Israel Deaconess Medical Center, Harvard Medical School, MA, USA,
    \num 15 \addr UNIDAD DE PATOLOGIA CLINICA:Guadalajara, JALISCO, Mexico,
    \num 16 \addr Department of Radiology, Children’s Health Orange County, CA, USA,
    \num 17 \addr Medical Artificial Intelligence (MAI) Lab, Crestview Radiology, Lagos, Nigeria,
    \num 18 \addr Sage Bionetworks, USA,
    \num 19 \addr Montreal Neurological Institute (MNI), McGill University, Montreal, QC, Canada,
    \num 20 \addr Department of Radiology, Perelman School of Medicine, University of Pennsylvania, Philadelphia, PA, USA,
    \num 21 \addr MLCommons, USA,
    \num 22 \addr Division of Computational Pathology, Department of Pathology and Laboratory Medicine, Indiana University School of Medicine, Indianapolis, IN, USA,
    \num 23 \addr Department of Neurosurgery at the University of Southern California, CA, USA,
    \num 24 \addr Department of Radiology, Duke University Medical Center, Durham, NC, USA,
    \num 25 \addr Mayo Clinic, MN, USA,
    \num 26 \addr Department of Informatics, Technical University Munich, Germany,
    \num 27 \addr National Cancer Institute, National Institutes of Health, Bethesda, MD, USA,
    \num 28 \addr Biomedical Engineering Rutgers University, New Brunswick, NJ, USA,
    \num 29 \addr Athinoula A Martinos Center for Biomedical Imaging, Massachusetts General Hospital and Harvard Medical School, Boston, MA, USA,
    \num 30 \addr Yale University, New Haven, CT, USA,
    \num 31 \addr PrecisionFDA, U.S. Food and Drug Administration, Silver Spring, MD, USA,
    \num 32 \addr Helmholtz AI, Helmholtz Munich, Germany,
    \num 33 \addr Department of Radiation Oncology, Duke University Medical Center, Durham, NC, USA,
    \num 34 \addr Department of Applied Mathematics and Computer Science, Technical University of Denmark, Denmark,
    \num 35 \addr Athinoula A Martinos Center for Biomedical Imaging, Massachusetts General Hospital, Boston, MA, USA,
    \num 36 \addr Booz Allen Hamilton, McLean, VA, USA,
    \num 37 \addr Biomedical Image Analysis \& Machine Learning, Department of Quantitative Biomedicine, University of Zurich, Switzerland,
    \num 38 \addr Mercy Catholic Medical Center, Darby, PA, USA,
    \num 39 \addr University of San Diego, CA, USA,
    \num 40 \addr Cornell University, Ithaca, NY, USA,
    \num 41 \addr Statistics in Imaging and Visualization Center, Department of Biostatistics, Epidemiology, and Informatics, University of Pennsylvania, Philadelphia, PA, USA,
    \num 42 \addr Department of Neuroradiology, Technical University of Munich, Munich, Germany,
    \num 43 \addr Greensboro Radiology, Greensboro, NC, USA,
    \num 44 \addr Center for AI and Data Science for Integrated Diagnostics (AI2D) \& Center for Biomedical Image Computing and Analytics (CBICA), University of Pennsylvania, Philadelphia, PA, USA,
    \num 45 \addr Brain Tumor Institute, Children's National Hospital, Washington, DC, USA,
    \num 46 \addr Brain Tumor Center, Cincinnati Children's Hospital, Cincinnati, OH, USA,
    \num 47 \addr Center for Cancer and Blood Disorders, Phoenix Children's Hospital, Phoenix, AZ, US,
    \num 48 \addr Center for Cancer and Blood Disorders, Children’s National Hospital, Washington, DC, USA
}

\abstract{
	Pediatric central nervous system tumors are the leading cause of cancer-related deaths in children. The five-year survival rate for high-grade glioma in children is less than 20\%. The development of    new treatments is dependent upon multi-institutional collaborative     clinical trials requiring reproducible and accurate centralized response assessment. We present the results of the BraTS-PEDs 2023 challenge, the first Brain Tumor Segmentation (BraTS) challenge focused on pediatric brain tumors. This challenge utilized data acquired from multiple international consortia dedicated to pediatric neuro-oncology and clinical trials. BraTS-PEDs 2023 aimed to evaluate volumetric segmentation algorithms for pediatric brain gliomas from magnetic resonance imaging using standardized quantitative performance evaluation metrics employed across the BraTS 2023 challenges. The top-performing AI approaches for pediatric tumor analysis included ensembles of nnU-Net and Swin UNETR, Auto3DSeg, or nnU-Net with a self-supervised framework. The BraTS-PEDs 2023 challenge fostered collaboration between clinicians (neuro-oncologists, neuroradiologists) and AI/imaging scientists, promoting faster data sharing and the development of automated volumetric analysis techniques. These advancements could significantly benefit clinical trials and improve the care of children with brain tumors.}

\keywords{Brain Tumor Segmentation, Pediatric Brain Tumors, BraTS-PEDs, Benchmarking}

\begin{document}

\maketitle\twocolumn

\section{Introduction}
	Although rare, pediatric tumors of the central nervous system are the leading cause of cancer-related death in children \citep{Ostrom2022, Hossain2021}. While they share some similarities with adult brain tumors, their imaging characteristics and clinical presentations differ significantly. For example, both adult glioblastoma (GBM) and pediatric diffuse midline glioma (DMG), including diffuse intrinsic pontine glioma (DIPG), are high-grade glial tumors with poor survival outcomes \citep{Mackay2017, Jansen2015}. However, whereas the incidence of adult GBM is 3 in 100,000 people, while DMG is approximately three times rarer. GBM primarily affects the frontal and temporal lobes in adults (median diagnosis age: 64 years) \citep{Thakkar2014}, but DMG predominantly arises in the pons and is diagnosed in children aged 5–10 years. Unlike GBM, which frequently presents with necrotic and enhancing regions on post-gadolinium T1-weighted MRI, DMG often lacks clear necrotic components at initial diagnosis. These differences underscore the need for dedicated imaging tools for pediatric brain tumors to improve diagnosis and prognosis \citep{Familiar2024}.

    Pediatric brain tumors exhibit high heterogeneity in histology and imaging features, often comprising subregions such as peritumoral edematous/infiltrated tissue, necrosis, and enhancing tumor margins, reflected in their radio-phenotypes on multi-parametric MRI (mpMRI) images \citep{Kazerooni2023a}. Tumors in surgically inaccessible locations, such as DIPG, rely primarily on longitudinal MRI for progression assessment. The current standard, defined by the Response Assessment in Pediatric Neuro-Oncology (RAPNO) cooperative working group  \citep{Cooney2020, Erker2020}, involves 2D linear measurements of tumor subregions on the axial slice of the largest extent \citep{Familiar2024}. However, manual segmentation is labor-intensive and prone to inter-operator variability, while 2D measurements provide an inaccurate surrogate for tumor volume due to tumor shape irregularities. Studies in adult and pediatric brain tumors have demonstrated the superiority of 3D volumetric measurements over 2D assessments in predicting clinical outcomes \citep{Ellingson2022, Lazow2022}.

    Automated tumor segmentation is critical for volumetric analysis, aiding in surgical planning, treatment response assessment, and longitudinal monitoring. The importance of volumetric tumor measurements—and, consequently, automated segmentation—has been increasingly recognized in pediatric neuro-oncology \citep{Jansen2015, Lazow2022}. Several automated tumor segmentation methods have been developed for pediatric brain tumors \citep{Kazerooni2023a, Vossough2024, Artzi2020, Kharaji2024, Liu2023, Mansoor2016, Tor-Diez2020, Peng2022, Nalepa2022, Gandhi2024, Boyd2024, Bareja2024, Hashmi2024}. Among available architectures, 3D U-Net \citep{Tor-Diez2020, Peng2022} and U-Net combined with ResNet \citep{Artzi2020} have been employed for whole tumor segmentation on T2-weighted or Fluid Attenuated Inversion Recovery (FLAIR) images and enhancing tumor segmentation on post-contrast T1-weighted imaging. The nnU-Net architecture \citep{Isensee2021} has emerged as a leading model for pediatric brain tumor segmentation, demonstrating strong performance across different pediatric tumor types \citep{Vossough2024, Kharaji2024, Liu2023, Gandhi2024, Bareja2024}.

    Transfer learning has also been applied, with models pretrained on adult glioma datasets and later fine-tuned for pediatric applications \citep{Kharaji2024, Liu2023, Nalepa2022, Boyd2024}. However, due to structural and developmental changes in the pediatric brain, transfer learning from adult to pediatric populations may not yield optimal results \citep{Boyd2024}. Beyond nnU-Net, alternative deep learning architectures have been explored, including MedNeXt, a ConvNeXt-based model optimized for brain tumor segmentation \citep{Hashmi2024}.

    Most segmentation studies before the Brain Tumor Segmentation in Pediatrics (BraTS-PEDs) challenge relied only on one or two MRI sequences \citep{Artzi2020, Tor-Diez2020, Peng2022, Nalepa2022, Boyd2024}. Few studies integrated multiparametric MRI, which is essential for accurately segmenting tumor subregions \citep{Kazerooni2023a, Vossough2024, Liu2023}. Furthermore, only a limited number of studies have developed pediatric-specific tumor subregion segmentation frameworks aligned with the RAPNO guidelines, spanning multiple histologies \citep{Kazerooni2023a, Vossough2024, Peng2022, Gandhi2024}. The absence of benchmarking platforms has further prevented standardized evaluation and comparison of these methods using a common validation dataset.

    The BraTS challenges, organized in conjunction with the Medical Image Computing and Computer Assisted Intervention (MICCAI, https://miccai.org) since 2012, have established a benchmarking framework for adult glioma segmentation \citep{Bakas2018, Menze2014, Bakas2017, Baid2021}. In 2023, the BraTS challenge expanded to include a variety of tumor entities such as adult glioma, intracranial meningioma  \citep{LaBella2023, LaBella2024a}, brain metastases \citep{Moawad2023}, Sub-Saharan African glioma, and pediatric tumors \citep{Kazerooni2023b} for the segmentation task. Additionally, new tasks were introduced, including Synthesis (Global) - Missing MRI \citep{Li2023}, Synthesis (Local) – Inpainting \citep{Kofler2023}, and Augmentation. The pediatric segmentation challenge (BraTS-PEDs) was initiated following the 2022 BraTS competition, where 60 pediatric DMGs were included in the test phase. The findings from this evaluation led to the expansion of the challenge in 2023, incorporating multi-consortia pediatric data for training, validation, and testing. 

    This manuscript presents the data, design, and outcomes of the BraTS-PEDs 2023 challenge, the first benchmarking initiative for pediatric brain tumor segmentation. By evaluating the performance of participating algorithms, this work aims to establish a standardized framework for pediatric brain tumor segmentation, facilitating clinical translation and extending applications to treatment response assessment in pediatric neuro-oncology.

\section{Data and Methods}
\subsection{Cohort Description}
	The BraTS-PEDs 2023 dataset      was supported by multiple consortia, including Children’s Brain Tumor Network (CBTN, https://cbtn.org/) \citep{Lilly2023}, International DIPG/DMG Registry (https://www.dipgregistry.org) and      COllaborative Network for NEuro-oncology Clinical Trials (CONNECT, https://connectconsortium.org/), along with additional clinical centers, including Boston's Children Hospital and Yale University. All cases contained pre- and post-gadolinium T1-weighted (labeled as T1 and T1CE, respectively), T2-weighted (T2), and T2-weighted fluid attenuated inversion recovery (FLAIR) images. These conventional mpMRI sequences are commonly acquired as part of standard clinical care for brain tumors. However, the image acquisition protocols, and MRI equipment differ across different institutions, resulting in heterogeneity in image quality and appearance in the multi-consortium cohort. Inclusion criteria comprised of pediatric subjects with: (1) histologically approved high-grade glioma, i.e., astrocytoma and DMG, including radiologically or histologically proven DIPG; (2) availability of all four structural mpMRI sequences on treatment-naive imaging sessions. Exclusion criteria included: (1) images that were of low quality or contained artifacts precluding reliable tumor segmentation, as determined by qualitative control (QC) by researchers experienced in pediatric brain tumor imaging and segmentation; and (2) infants younger than one month of age. The latter criteria was imposed as the neonatal brain within the first month undergoes rapid developmental changes, including immature myelination and evolving tissue contrasts \citep{Gilmore2007}, which can complicate image segmentation and analysis.  In the 2023 BraTS-PEDs, data for 167 patients were included (CBTN, n = 113; Boston's Children Hospital, n = 30; Yale University, n = 24). Data from the International DIPG/DMB Registry was only processed but not segmented for the 2023 challenge, but will be included in the 2024 challenge.
    
\subsection{Imaging Data}
    For all patients, mpMRI scans were pre-processed using a standardized approach, including conversion of the DICOM files to the NIfTI file format, co-registration to the same anatomical template (i.e., SRI24) \citep{Rohlfing2010}, and resampling to an isotropic resolution of 1 mm3. The pre-processing pipeline is publicly available through the Cancer Imaging Phenomics Toolkit (CaPTk) \citep{Pati2020, Rathore2018, Davatzikos2018}, Federated Tumor Segmentation (FeTS) tool https://github.com/FETS-AI/Front-End/ \citep{Pati2022}, and the docker container for pediatric-specific pipeline including skull-stripping and tumor segmentation on GitHub: https://github.com/d3b-center/peds-brain-seg-pipeline-public. De-identification was performed through removing protected health information from DICOM headers. Defacing was performed via skull-stripping using a pediatric-specific skull-stripping method \citep{Kazerooni2023a} to prevent any potential facial reconstruction/recognition of the patients. Figure 1A presents the data preparation workflow.

    The pre-processed images were segmented into tumor subregions using either of two pediatric automated deep learning segmentation models \citep{Kazerooni2023a, Liu2023} before being manually corrected by pediatric neuro-radiologists. The tumors were segmented into four subregions. These subregions comprising of enhancing tumor (ET), non-enhancing tumor (NET), cystic component (CC), and peritumoral edema (ED) regions, with definitions provided in \citep{Familiar2024}, and summarized below:
    \begin{itemize}
    \item ET is described by areas with enhancement (brightness) on T1 post-contrast images as compared to T1 pre-contrast. In case of mild enhancement, checking the signal intensity of normal brain structures can be helpful.
    \item CC typically appears with hyperintense signal (very bright) on T2 and hypointense signal (dark) on T1CE. The cystic portion should be within the tumor, either centrally or peripherally (as compared to ED which is peritumoral). The brightness of CC is defined as comparable or close to cerebrospinal fluid (CSF).
    \item NET is defined as any other abnormal signal intensity within the tumorous region that cannot be defined as enhancing or cystic. For example, the abnormal signal intensity on T1, FLAIR, and T2 that is not enhancing on T1CE should be considered as non-enhancing portion.
    \item ED is defined by the abnormal hyperintense signal (very bright) on FLAIR scans. ED is finger-like spreading that preserves underlying brain structure and surrounds the tumor.
    \end{itemize}

    The American Society of Neuroradiology collaborated in generating ground truth annotation for the majority of data in this challenge (ASNR, https://www.asnr.org/). Four labels were generated (Figure 1) using the preliminary automated segmentation and were then manually revised by ASNR volunteer neuroradiology experts of varying rank and experience in accordance with the annotation guidelines. The expert annotators were provided with the four mpMRI sequences (T1, T1CE, T2, FLAIR) along with the fused automated segmentation volume to initiate the manual refinements using the ITK-SNAP software \citep{Yushkevich2006}. After segmentation corrections by the annotators, three attending board-certified neuroradiologists reviewed the segmentations to either approve or return to the individual annotators for further refinements. This process was followed iteratively until the approvers found the refined tumor subregion segmentations acceptable for public release and the challenge conduction. The final segmentations were provided and used as ground truth for model training and evaluation.

    \begin{figure*}[htb]
		\centering
		\includegraphics[width=0.75\linewidth, height=.5\textheight]{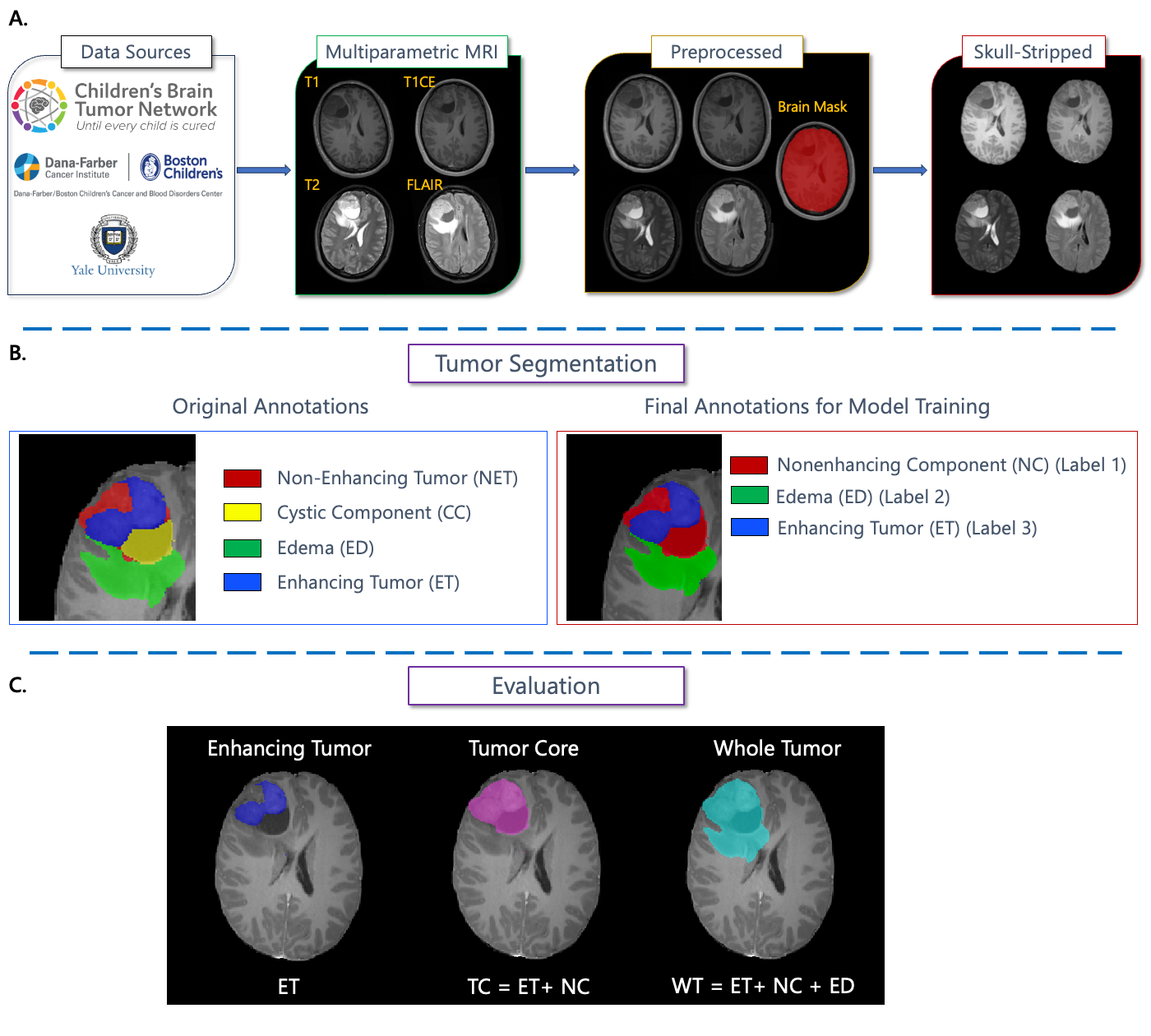}
		\caption{A schematic diagram of different steps in data preparation for BraTS-PEDs 2023 data: (A) Data preparation process; (B) Tumor segmentation approach: Left - original annotations including four subregions, Right – final annotations provided to the teams for model training (the final annotations and their label values were prepared in agreement with other BraTS 2023 challenges). The color maps in the images shown in Figure 1B illustrate the label descriptions provided in the original annotations given to the annotators (left), as compared to the final annotations (right) – which adhere to the standards of other BraTS challenges – provided to the teams; (C) Evaluation labels: for calculating the performance metrics, ET, TC (combination of ET and NC), and WT (a combination of ET, NC, ED) were used.}
	\end{figure*}

    Although four tumor subregions were annotated for all the pediatric data, to follow the guidelines of the other BraTS 2023 challenges and to allow the teams to potentially incorporate data from adult glioma challenge into their model training, the teams were provided with only three segmentation labels, i.e., non-enhancing component (NC – the union of non-enhancing tumor, cystic component, and necrosis; label 1), ED (label 2), and ET (label 3), as shown in Figure 1B. The color maps in the images shown in Figure 1B illustrate the label descriptions provided in the original annotations given to the annotators (Figure 1B; left), as compared to the final annotations (Figure 1B; right) – which adhere to the standards of other BraTS challenges – provided to the teams.

\subsection{Challenge Setup}
    The BraTS-PEDs 2023 cohort was split into training (n = 98), validation (n = 44), and testing datasets (n = 24). Training and validation included a random split between CBTN and Boston cases. All Yale cases were used for independent testing. The training data shared with the teams comprised mpMRI scans and ground truth labels for the training cohort, released in mid-May 2023. The cohort shared for validation included only the mpMRI sequences without associated ground truth and it was released in late June 2023. Notably, the testing cohort was withheld from the participants.

    Teams were prohibited from training their algorithm on any additional public and/or private data besides the provided BraTS-PEDs 2023 data or to using models pretrained on other datasets. This restriction was imposed to allow for a direct and fair comparison among the participating methods. However, teams were allowed to use additional public and/or private data for publication of their scientific papers, on the condition that they also provide a report of their results on the data from BraTS-PEDs 2023 challenge alone and discuss potential differences in the obtained results.

    The methods submitted to the BraTS-PEDs 2023 challenge were evaluated based on performances for the segmentation of ET, tumor core (TC – the combination of ET and NC), and the entire tumorous region, or whole tumor (WT – the union of ET, NC, and ED) (Figure 1C). The containerized submissions of the teams were evaluated on the synapse.org platform supported by Medperf \citep{Karargyris2023}. Teams were required to send the output of their methods to the evaluation platform for the scoring to occur during the training and the validation phases. At the end of the validation phase, teams were asked to identify the method they would like to be evaluated by the organizers in the final testing/ranking phase. The training, validation, and testing phases ran between May-July 2023, July-August 2023, and August 2023, respectively. The results and ranking of the challenge were presented at the MICCAI 2023 conference on October 10, 2023.

\subsection{Benchmark Design: Evaluation and Ranking}
    The evaluation and ranking were based on a lesionwise ranking scheme in which the ranking for each team was computed relative to other challenge competitors for each of the subjects in the test set, for each tumor region, e.g., ET, TC, and WT. We used two established performance evaluation metrics, i.e., Dice similarity coefficient (Dice) and 95\% Hausdorff distance (HD95) using the Generally Nuanced Deep Learning Framework (GaNDLF; gandlf.org) \citep{Pati2023}. At BraTS-PEDs 2023, each team was ranked for 24 test subjects, 3 tumor regions, and 2 metrics, resulting in 144 individual rankings. Then for each team, the final ranking score (FRS) was computed by producing a cumulative rank as an average of all individual rankings for each patient, followed by averaging the cumulative ranks across all testing subjects for each competing team. 

    The lesion-wise Dice and HD95 were calculated by isolating and identifying each disjoint lesion, through dilating the ground truth segmentation masks for each tumor region, i.e., ET, TC, and WT, by 3 pixels in all directions (i.e., by a size of 3x3x3). Dice and HD95 scores were calculated for each component or lesion individually, after penalizing all false positive and false negative values with a score of 0. Subsequently, the mean value of Dice and HD95 scores across all lesions were calculated for each subject. The lesions below a cutoff value of 50 voxels were removed from any further evaluation. If the models were not able to predict a region for a subject, the Dice score was assigned a value of 0 and HD95 was set to 374 (the distance between corners in the SRI atlas space), and if they correctly did not predict a non-existing region, the Dice score and HD95 were given values of 1 and 0, respectively. Lesion-wise Dice and HD95 scores for each case were calculated according to equations 1 and 2. More information on the evaluation metrics along with the codes can be found at https://github.com/rachitsaluja/BraTS-2023-Metrics.

    \begin{equation}
    \text{Lesionwise Dice Score} = \frac{\Sigma_{i}^{L}\text{Dice}(l_{i})}{\text{TP}+\text{FN}+\text{FP}}
    \end{equation}
    \begin{equation}
    \text{Lesionwise HD95} = \frac{\Sigma_{i}^{L}\text{HD95}(l_{i})}{\text{TP}+\text{FN}+\text{FP}}
    \end{equation}

    Where L represents the number of ground truth lesions, TP denotes the number of true positives, FN indicates the number of false negatives, and FP signifies the number of false positives.

    Finally, the statistical differences in performance, beyond those expected by chance, between each pair of teams were determined using permutation testing. This was achieved for each pair of teams, by randomly permuting the cumulative ranks for each subject and calculating the difference in FRS between the pair of teams in each permutation. Statistical significance of rankings was determined using a p-value when the proportion of times the difference in FRS using the permutated data exceeded the observed difference in FRS using the actual data.  These values are reported in an upper triangular matrix.

\subsection{Data Availability}
    The training data, including ground truth segmentations, and the validation data without ground truth segmentations can be accessed and downloaded from the Synapse  www.synapse.org/Synapse:syn51156910/wiki/627000. More information about the BraTS-PEDs 2023 challenge is at www.synapse.org/Synapse:syn51156910/wiki/622461.

\section{Results} 
    \subsection{Overview of the Submissions}
        Among the teams that participated in the training phase, 18 teams submitted a total of 159 entries during the validation phase, with 147 of these submissions receiving scores. Nine teams completed all challenge stages and were evaluated during the testing phase of the BraTS‐PEDs 2023 challenge \citep{Capellán-Martín2023, Myronenko2023, Zhou2023, Huang2023, Maani2023, Mistry2023, Javaji2023, Mantha2023,Temtam2023}. During the competition, participants were allowed unlimited validation of their algorithms on the Synapse platform using validation data with withheld ground truth, which enabled them to iteratively fine-tune their models before making their final submission for testing. The validation outcomes for each team is reported in their respective publications \citep{Capellán-Martín2023, Myronenko2023, Zhou2023, Huang2023, Maani2023, Mistry2023, Javaji2023, Mantha2023,Temtam2023}.
        
    \subsection{Summary of Participating Methods in BraTS-PEDs 2023 Challenge}
        The participating teams in the BraTS-PEDs 2023 challenge utilized diverse deep learning-based segmentation approaches, incorporating various architectures, training strategies, and post-processing techniques to tackle the complexities of pediatric brain tumor segmentation. Below, we provide a summarized overview of these methods (see Table 1 for a summary of the proposed techniques by each team), with further details available in the individual publications  \citep{Capellán-Martín2023, Myronenko2023, Zhou2023, Huang2023, Maani2023, Mistry2023, Javaji2023, Mantha2023,Temtam2023}.

        Several teams used variants of the U-Net architecture, with modifications to improve feature extraction and segmentation accuracy. The BiomedMBZ team implemented SegResNet and MedNext with deep supervision and extensive pre- and post-processing \citep{Maani2023}. Blackbean introduced the Scalable and Transferable U-Net (STU-Net), a large-scale model with up to 1.4 billion parameters, pre-trained on the TotalSegmentator dataset \citep{Huang2023}. 

        Several teams adopted ensemble approaches to improve robustness. CNMCPMMI combined nnU-Net with Swin UNETR, applying label-wise ensembling and cross-validated thresholding to refine segmentation outputs \citep{Capellán-Martín2023}. SchlaugLab developed a model combining ONet and modified U-Net variants, integrating custom loss functions and composite data augmentation \citep{Javaji2023}. SherlockZyb extended nnU-Net with self-supervised pre-training and an adaptive region-specific loss function to handle tumor heterogeneity and class imbalances \citep{Zhou2023}.

        Automated model selection was explored by NVAUTO, which utilized Auto3DSeg from MONAI, requiring minimal user input while training SegResNet with 5-fold cross-validation and deep supervision \citep{Myronenko2023}. Meanwhile, UMNIverse introduced a GAN-based model, Temporal Cubic PatchGAN (TCuP-GAN), incorporating Convolutional Long Short-Term Memory Networks (ConvLSTMs) to leverage temporal MRI features \citep{Mantha2023}. Other teams focused on novel architectural modifications. Isahajmistry team enhanced nnU-Net with Omni-Dimensional Dynamic Convolution (ODConv3D) and multi-scale attention mechanisms for improved feature learning \citep{Mistry2023}. VisionLab proposed a Multiresolution Fractal Deep Neural Network (MFDNN), using wavelet-based convolution layers to extract tumor textures at multiple scales while performing uncertainty analysis to enhance segmentation reliability \citep{Temtam2023}.

    \subsection{Evaluation and Ranking Results}
        Table 1 provides a summary of the cumulative ranks and FRS for each team across all test subjects, including the teams' rankings, and a high-level description of the methods used by each team. The CNMCPMI2023 team \citep{Capellán-Martín2023} achieved the highest score among the teams, followed by NVAUTO \citep{Myronenko2023}, SherlockZyb \citep{Zhou2023}, and Blackbean \citep{Huang2023} (see Table 1 that illustrates the scores and methods used by all teams that completed the challenge tasks).

        Figure 2 illustrates the p-values calculated between pairs of teams to evaluate statistical significance between the achieved performance. Specifically, among the best performing teams, the results for CNMCPMI2023 and NVAUTO (p = 0.45), NVAUTO and SherlockZyb (p = 0.1), and SherlockZyb and Blackbean (p = 0.43) were not significantly different. This may be due to the small sample size of the test cohort (n = 24) for this challenge.

        Figures 3 to 5 display the Dice and HD95 scores for all test subjects for all the participating teams. Tables 2 to 4 summarize the performance metrics—Dice, HD95, and sensitivity—of each team. The CNMCPMI2023 $|$ NVAUTO $|$ SherlockZyb $|$ Blackbean teams achieved Dice scores of 0.83 $|$ 0.84 $|$ 0.83 $|$ 0.81 for WT, 0.81 $|$ 0.78 $|$ 0.77 $|$ 0.79 for TC, and 0.65 $|$ 0.55 $|$ 0.63 $|$ 0.53 for ET segmentation, and HD95 values of 20.86 $|$ 18.05 $|$ 6.11 $|$ 23.56 for WT, 21.82 $|$ 27.10 $|$ 22.28 $|$ 22.02 for TC, and 43.89 $|$ 115.32 $|$ 59.68 $|$ 22.02 for ET segmentation. Interestingly, the sensitivity of the NVAUTO algorithm for the segmentation of WT and ET was higher than that of the other two teams, indicating a higher overall rate of true positive predictions by this algorithm. However, this increased sensitivity came at the cost of over-segmentation, resulting in more false positives and less accurate boundary placement for ET segmentation.

        The prediction results for ET demonstrate a greater variability and skewness in the Dice score distribution, as indicated by the mean and median values in Table 2 and the box plots in Figure 3. In contrast, the segmentation of TC and WT (Figures 4-5) shows less variability and skewness. Given the small volume or absence of the ET region in DMG tumors, the Dice score more heavily penalizes segmentation errors in the ET region. TC and WT segmentations are more consistent and accurate across all teams, with WT segmentation being the most precise. This indicates that, on average, most methods reliably locate the lesion within MRI scans, while distinguishing between different tumor compartments is more challenging.

        Figure 6 showcases examples of the performances of CNMCPMI2023, NVAUTO, SherlockZyb, and Blackbean teams for tumor segmentation in four testing subjects. Panels (A-B) indicate the best results for these three teams, with WT Dice above 0.90. Panels (C-D) present the subjects with worst results across the top performing teams. As indicated in these examples, there are situations where one algorithm performs better than the remaining algorithms in segmentation of tumorous regions. In Figure 6(C), CNMCPMI2023, SherlockZyb, and Blackbean achieved Dice scores of 0.88, while NVAUTO achieved a low Dice score (0.18) for WT segmentation. This low performance for NVAUTO was achieved due to many false positive results for segmentation of the tumor for this subject. On the other hand, in the example presented in Figure 6(D), NVAUTO and Blackbean algorithms produced WT segmentation results with Dice scores of 0.60 and 0.70, respectively, while the other two algorithms, i.e., CNMCPMI2023 and SherlockZyb, did not generate any segmentation outputs (WT Dice = 0 or 0.05). This situation underscores the potential benefit of ensembling different algorithms in achieving better overall performance.

    \begin{table*}[h!] 
        \begin{minipage}{\textwidth}
        \centering
		\caption{Cumulative ranks across testing subjects and final ranking score (FRS) for the competing teams in the BraTS-PEDs2023 challenge. The lower ranks show better performance.}
        \resizebox{\textwidth}{!}{
		\begin{tabular}{lcccc}
			\textbf{Teams} & \textbf{Method} & \textbf{Cumulative Ranks Across Subjects} & \textbf{FRS} & \textbf{Rank in the Challenge} \\
			\hline
			CNMCPMI2023 & Ensemble of nnU-Net and Swin UNETR  & 241.5 & 10.06 & - \footnote{Participating teams that included challenge organizers underwent evaluation without ranking to uphold fairness in our benchmarking initiative. This approach aligns with MICCAI guidelines, as well as the CBTN policy for hackathons.} \\
			NVAUTO & Auto3DSeg & 246 & 10.25 & 1 \\
            SherlockZyb & nnU-Net (self-supervised) & 286.5 & 11.93 & 2 \\
            Blackbean & Scalable and Transferable U-Net (STU-Net) & 290.5 & 12.1 & 3 \\
            BiomedMBZ & Ensemble of SegResnet and MedNext & 318 & 13.25 & 4 \\
            SchlaugLab & Ensemble of ONet and UNet & 395.5 & 16.47 & 5 \\
            VisionLabODU23 & Multiresolution Fractal Deep Neural Network (MFDNN) & 427 & 17.79 & 6 \\
            Isahajmistry & Modified nnU-Net & 483 & 20.12 & 7 \\
            UMNiverse & Temporal Cubic PatchGAN (TCuP-GAN) & 552 & 23 & 8 
		\end{tabular}
        }
    \end{minipage}
	\end{table*}

    \begin{figure*}[h!]
		\centering
		\includegraphics[width=0.8\linewidth, height=.5\textheight]{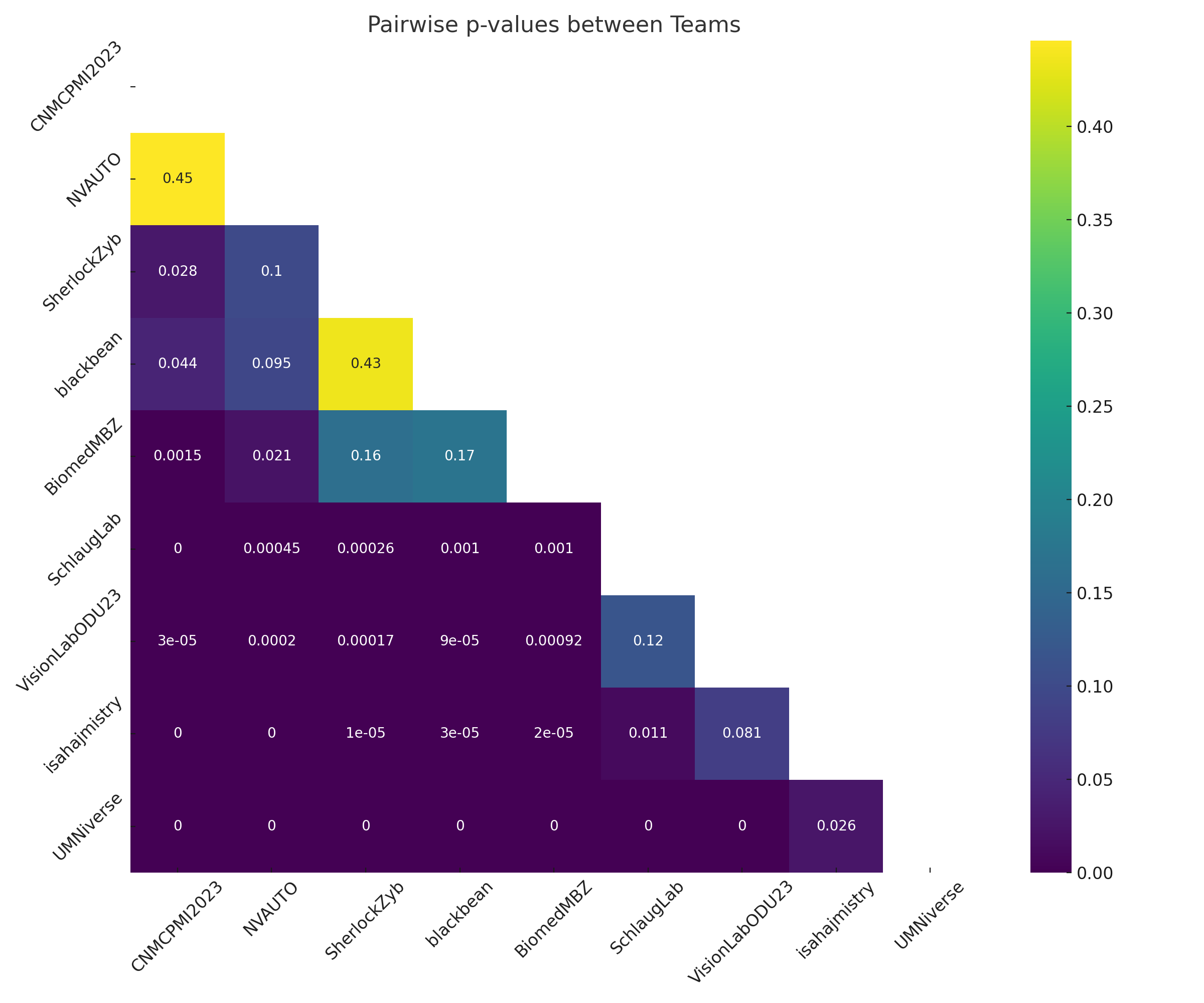}
		\caption{Pairwise p-values between the participating teams.}
	\end{figure*}

    \begin{figure*}[h!]
		\centering
		\includegraphics[width=0.9\linewidth, height=.4\textheight]{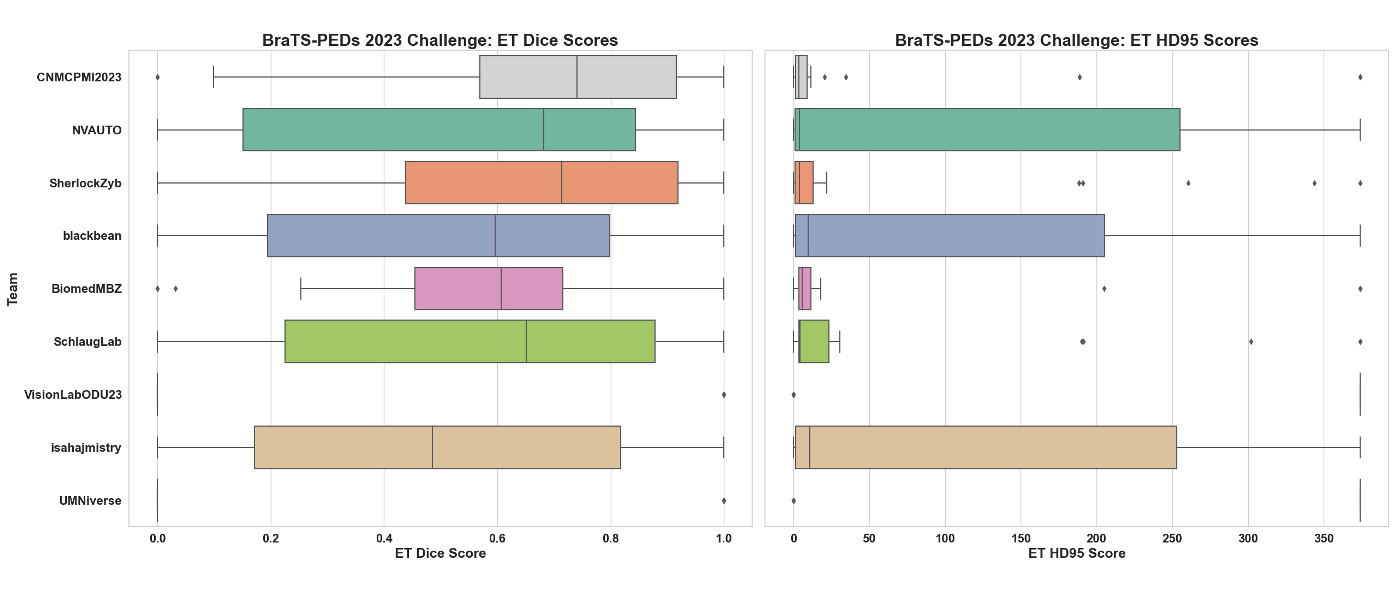}
		\caption{Dice and HD95 Scores for segmentation of enhancing tumor (ET) across participating teams in the BraTS-PEDs 2023 Challenge.}
	\end{figure*}

    \begin{figure*}[h!]
		\centering
		\includegraphics[width=1.0\linewidth, height=.4\textheight]{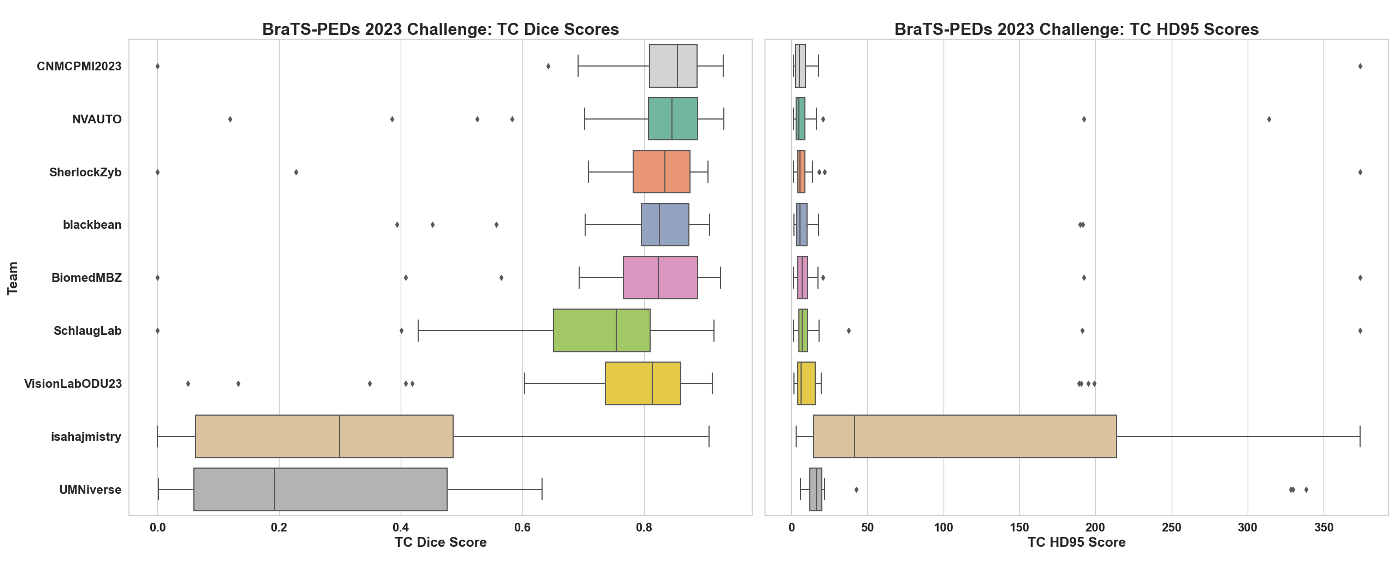}
		\caption{Dice and HD95 Scores for segmentation of tumor core (TC) across participating teams in the BraTS-PEDs 2023 Challenge.}
	\end{figure*}

    \begin{figure*}[h!]
		\centering
		\includegraphics[width=1.0\linewidth, height=.4\textheight]{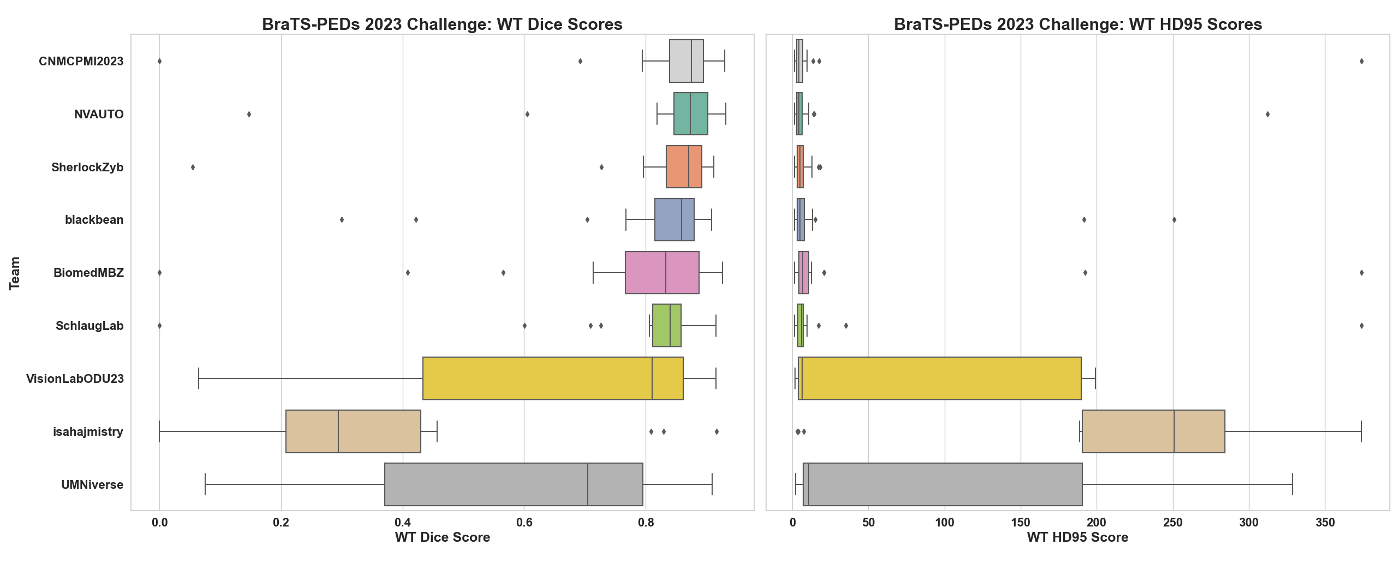}
		\caption{Dice and HD95 Scores for segmentation of whole tumor (WT) across participating teams in the BraTS-PEDs 2023 Challenge.}
	\end{figure*}

    \begin{table*}[h!] 
        \centering
		\caption{Summary of the performance evaluation metrics for segmentation of enhancing tumor (ET) region for the participating teams. The results are shown in mean $\pm$ standard deviation (median) format.}
		\begin{tabular}{lccc}
			\textbf{Teams} & \textbf{Dice Similarity Coefficient} & \textbf{95\% Hausdorff Distance} & \textbf{Sensitivity} \\
			\hline
			CNMCPMI2023 & 0.65 ± 0.32 (0.74) & 43.89 ± 108.59 (3.67)  & 0.70 ± 0.18 (0.74) \\
			NVAUTO & 0.55 ± 0.37 (0.68) & 115.32 ± 144.03 (4.04) & 0.77 ± 0.20 (0.80) \\
            SherlockZyb & 0.63 ± 0.32 (0.71) & 59.68 ± 116.70 (3.93) & 0.71 ± 0.27 (0.74) \\
            Blackbean & 0.53 ± 0.34 (0.60) & 22.02 ± 52.12 (5.83) & 0.63 ± 0.31 (0.71) \\
            BiomedMBZ & 0.55 ± 0.27 (0.61) & 45.32 ± 109.10 (5.69) & 0.80 ± 0.29 (0.92) \\
            SchlaugLab & 0.55 ± 0.36 (0.65) & 65.23 ± 122.23 (4.30) & 0.55 ± 0.37 (0.59) \\
            VisionLabODU23 & 0.17 ± 0.38 (0) & 311.67 ± 142.38 (374) & 0.17 ± 0.38 (0) \\
            Isahajmistry & 0.48 ± 0.36 (0.49) & 128.12 ± 147.11 (10.85) & 0.66 ± 0.35 (0.82) \\
            UMNiverse & 0.17 ± 0.38 (0) & 311.67 ± 142.38 (374) & 0.17 ± 0.38 (0)  
		\end{tabular}
	\end{table*}

    \begin{table*}[h!] 
        \centering
		\caption{Summary of the performance evaluation metrics for segmentation of tumor core (TC) region for the participating teams. The results are shown in mean $\pm$ standard deviation (median) format.}
		\begin{tabular}{lccc}
			\textbf{Teams} & \textbf{Dice Similarity Coefficient} & \textbf{95\% Hausdorff Distance} & \textbf{Sensitivity} \\
			\hline
			CNMCPMI2023 & 0.81 ± 0.18 (0.85) & 21.82 ± 75.15 (5.26)   & 0.76 ± 0.18 (0.81)\\
			NVAUTO & 0.78 ± 0.19 (0.85) & 27.10 ± 72.11 (5.08) & 0.74 ± 0.14 (0.77) \\
            SherlockZyb & 0.77 ± 0.21 (0.83) & 22.28 ± 75.08 (5.70) & 0.68 ± 0.21 (0.72) \\
            Blackbean & 0.79 ± 0.14 (0.83) & 22.02 ± 52.15 (5.83) & 0.75 ± 0.11 (0.79) \\
            BiomedMBZ & 0.77 ± 0.20 (0.82) & 30.36 ± 82.47 (7.21) & 0.73 ± 0.19 (0.92) \\
            SchlaugLab & 0.70 ± 0.20 (0.75) & 31.61 ± 82.20 (7.00) & 0.61 ± 0.20 (0.65) \\
            VisionLabODU23 & 0.71 ± 0.25 (0.81) & 38.40 ± 71.01 (6.5) & 0.70 ± 0.23 (0.77) \\
            Isahajmistry & 0.32 ± 0.29 (0.30) & 130.25 ± 135.52 (41.59) & 0.28 ± 0.25 (0.21) \\
            UMNiverse & 0.25 ± 0.22 (0.19) & 55.15 ± 107.21 (16.35) & 0.19 ± 0.16 (0.14)  
		\end{tabular}
	\end{table*}

    \begin{table*}[h] 
        \centering
		\caption{Summary of the performance evaluation metrics for segmentation of whole tumor (WT) region for the participating teams. The results are shown in mean $\pm$ standard deviation (median) format.}
		\begin{tabular}{lccc}
			\textbf{Teams} & \textbf{Dice Similarity Coefficient} & \textbf{95\% Hausdorff Distance} & \textbf{Sensitivity} \\
			\hline
			CNMCPMI2023 & 0.83 ± 0.18 (0.87) & 20.86 ± 75.31 (4.24)   & 0.76 ± 0.18 (0.81)\\
			NVAUTO & 0.84 ± 0.16 (0.87) & 18.05 ± 62.77 (4.30) & 0.80 ± 0.09 (0.82) \\
            SherlockZyb & 0.83 ± 0.17 (0.87) & 6.11 ± 4.50 (4.79) & 0.75 ± 0.17 (0.80) \\
            Blackbean & 0.81 ± 0.15 (0.86) & 23.56 ± 61.60 (4.95) & 0.78 ± 0.07 (0.80) \\
            BiomedMBZ & 0.78 ± 0.20 (0.82) & 30.45 ± 82.46 (6.85) & 0.72 ± 0.19 (0.77) \\
            SchlaugLab & 0.79 ± 0.18 (0.84) & 22.36 ± 75.20 (5.92) & 0.71 ± 0.18 (0.75) \\
            VisionLabODU23 & 0.68 ± 0.26 (0.81) & 68.18 ± 69.56 (6.35) & 0.73 ± 0.21 (0.79) \\
            Isahajmistry & 0.35 ± 0.23 (0.29) & 221.95 ± 99.53 (250.67) & 0.76 ± 0.20 (0.81) \\
            UMNiverse & 0.60 ± 0.25 (0.70) & 72.62 ± 107.48 (10.69) & 0.61 ± 0.16 (0.59)  
		\end{tabular}
	\end{table*}

    \begin{figure*}[h]
		\centering
		\includegraphics[width=1.0\linewidth, height=.45\textheight]{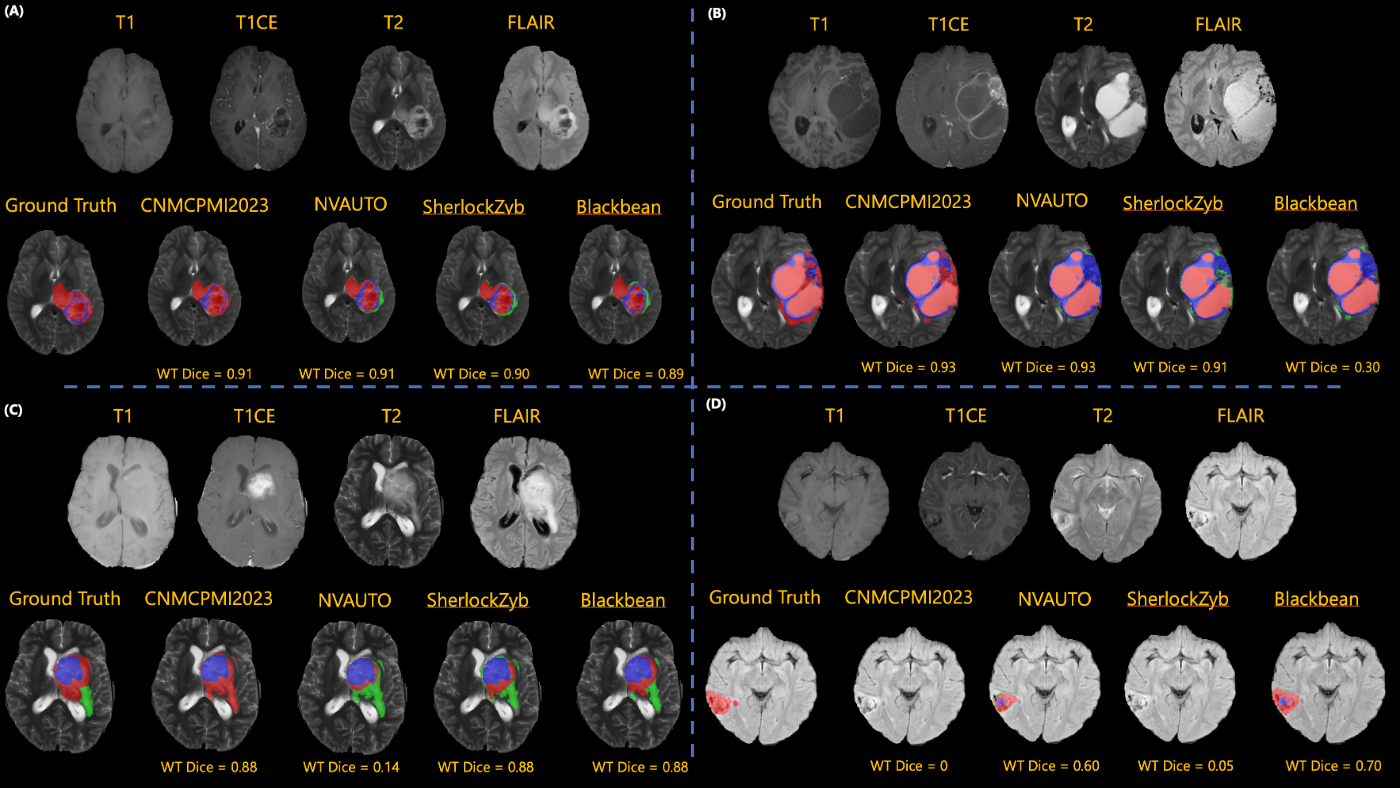}
		\caption{Examples of tumor segmentation for sample testing patients across the best performing teams: (A-B) best cases; (C-D) worst cases.}
	\end{figure*}   

\section{Discussion}
    The results of the BraTS-PEDs 2023 challenge provide valuable insights into different automated volumetric segmentation methods for pediatric brain tumors. The comprehensive evaluation metrics and rigorous validation phase in this challenge highlight the strengths and weaknesses of the various methods, offering a clear benchmark for future developments.
    \subsection{Overview of the Challenge Results}
        In 2022, the BraTS Glioma challenge tested top-performing methods—developed for adult glioma segmentation—on a small set of unseen pediatric brain tumor data. Results revealed that although models trained on adult brain tumors achieved high Dice scores for segmenting the whole pediatric tumor from mpMRI scans, their performance on tumor subregions (such as ET or NC) was inaccurate. This confirmed our hypothesis that effective pediatric segmentation techniques require training on a curated and annotated pediatric brain tumor dataset. However, as with any machine learning and deep learning endeavor, developing a model that generalizes well necessitates a large, standardized dataset. To address this, the BraTS-PEDs 2023 initiative has compiled the most extensive annotated collection of pediatric high-grade gliomas, including astrocytomas and DMGs.

        The top-performing methods at BraTS-PEDs 2023 included an ensemble method combining nnU-Net and Swin UNETR, Auto3DSeg using the SegResNet algorithm, and an extension of the nnU-Net model with self-supervised pretraining integrated with adaptive region-specific loss. All of these methods are based on the U-Net convolutional neural network architecture. The models achieved mean Dice scores of 0.83-0.84 for WT segmentation and slightly lower yet respectable Dice scores of 0.77-0.81 for TC segmentation. However, lower and more variable performance was observed in the segmentation of the ET region, with mean Dice scores of 0.65, 0.63 and 0.55 across the three top performing teams, which emphasizes the unique challenges posed by pediatric brain tumors. The small volume or absence of ET region in DMG likely contributes to the greater variability and skewness in the Dice score distribution for this subregion. This finding indicates that while most methods can reliably locate the WT within MRI scans, accurately distinguishing between different tumor compartments remains a significant challenge in pediatric brain tumors. 

        The examples in the manuscript demonstrate that no single algorithm consistently outperforms others across all cases. The variability in performance, especially in challenging scenarios, highlights the need for continued research and development in this field. Specifically, ensembling the top-performing models may achieve better overall performance for segmentation of multi-institutional data by leveraging their individual strengths and mitigating      their weaknesses. For instance, a method with high sensitivity, despite relatively lower Dice and higher HD95 scores, can be useful when it is critical not to miss any positive regions (such as in ET segmentation). By combining this method with others, it can be refined to reduce false positives and improve boundary accuracy.

    \subsection{Clinical Implications}
        The advancements in segmentation algorithms, as showcased by the BraTS-PEDs 2023 challenge, have several clinical implications. Detailed automated segmentations can enable radiation oncologists to better target treatment areas while sparing healthy tissue, ultimately reducing side effects \citep{Bibault2024}. Furthermore, the integration of such automated tools into clinical workflows may expedite diagnostic processes and reduce inter-observer variability, facilitating more standardized and timely treatment decisions. Importantly, consistent monitoring facilitated by these algorithms could lead to prompt interventions, which is particularly crucial in pediatric populations where timely treatment is essential to reduce long-term adverse developmental impacts. Similarly, accurate tumor delineation can improve surgical planning by providing neurosurgeons with precise maps of tumor boundaries, thereby improving resection strategies and minimizing damage to critical brain structures \citep{Sulangi2024}. Future work should focus on prospective clinical validations to assess the real-world benefits and cost-effectiveness, ensuring they contribute meaningfully to patient outcomes and personalized treatment strategies.

    \subsection{Limitations}
        Unlike previous challenges (e.g., BRATS \citep{Baid2021, Moawad2023, LaBella2024b}) that focused on adult populations, the BraTS‐PEDs 2023 challenge specifically addressed brain tumor segmentation in pediatric patients, who present unique challenges such as age-related anatomical variations. Nonetheless, our analysis reaffirms common limitations of current deep learning paradigms, including limited generalizability across diverse acquisition protocols and dependence on large, well-annotated datasets, that continue to impede robust clinical translation. 

        BraTS-PEDs 2023 also faced specific limitations, including a small dataset size and a narrow range of pediatric brain tumor histologies. For instance, the statistical analysis comparing the top teams revealed no significant performance differences, which may be attributed to the small sample size of the test cohort (n = 24). Future challenges would benefit from larger test cohorts to better discern performance differences and validate algorithm robustness.

        In addition, to align with the rest of the BraTS 2023 cluster of challenges, our evaluation focused on segmenting the enhancing tumor, tumor core, and whole tumor. This focus may limit applicability of these methods for response assessment in pediatric brain tumors. Recognizing the distinct differences between adult and pediatric brain tumors \citep{Mackay2017}, especially in components critical for response assessment \citep{Mackay2017, Thakkar2014, Familiar2024}, we will customize future evaluations to better address pediatric tumor characteristics. Furthermore, the typically small size of ET regions in DMG tumors prompted us to set a threshold of 50 voxels for the detection of a true positive, which might have impacted the accuracy of ET predictions. Moving forward, we intend to adopt a more data-driven approach to setting this threshold in subsequent challenges.

    \subsection{Future Directions}
        Encouraged by the results of the BraTS-PEDs 2023 challenge, our next steps include expanding the dataset to include subjects from additional institutions and incorporating various histologies of high- and low-grade pediatric glioma. This initiative will offer the research community access to a comprehensive dataset of rare pediatric tumors with curated mpMRI and annotation, thereby aiding in the development of advanced tools for computer-assisted treatment planning and prognosis. Future data collection will also encompass post-operative and post-treatment scans and clinical outcomes. The data is publicly available at www.synapse.org/Synapse:syn51156910/wiki/627000.

\section{Conclusion}
The BraTS-PEDs 2023 challenge provided an open-access, curated, annotated dataset of multi-sequence MRIs of pediatric brain tumors. The challenge also provided the platform to evaluate methods for automated volumetric segmentation of pediatric brain tumors, thereby supporting the development of techniques that enhance decision support systems for assessing treatment responses and predicting the outcomes of these rare conditions. This manuscript presented the overview and results of the BraTS-PEDs 2023 challenge.


\acks{Success in any challenging medical domain depends upon the quality of well annotated multi-institutional datasets. We are grateful to all the data contributors, annotators, and approvers for their time and efforts. Our profound thanks go to the Children’s Brain Tumor Network (CBTN), the Collaborative Network for Neuro-oncology Clinical Trials (CONNECT), the International DIPG/DMG Registry (IDIPGR), the American Society of Neuroradiology (ASNR), and the Medical Image Computing and Computer Assisted Intervention (MICCAI) Society for their invaluable support of this challenge.

Research reported in this publication was partly supported by the National Institutes of Health (NIH) award: NCI/ITCR:U01CA242871 and NCI:UH3CA236536, and by grant funding from the Pediatric Brain Tumor Foundation and DIPG/DMG Research Funding Alliance (DDRFA). The content of this publication is solely the responsibility of the authors and does not represent the official views of the NIH.}

%
\ethics{For this retrospective study, informed consent had been obtained from all subjects at their respective institutions or a waiver of informed consent was approved by the local institutional review board. The protocol for releasing the data was approved by the institutional review board of the data-contributing institution. All patients were fully de-identified and stripped of any patient identifiers before sharing with the organizing team. Further de-identification was performed through renaming all images and the respective ground truth segmentations, and  defacing all MRI images through skull-stripping before sharing with the participants.}

\coi{None of the authors have any conflicts of interest related to this manuscript.}

\bibliography{sample}


\end{document}